\documentclass[journal=jpccck,manuscript=letter,layout=twocolumn]{achemso}
\usepackage{multirow}
\usepackage{chemformula} 
\usepackage[T1]{fontenc} 
\usepackage{hyperref}
\usepackage{comment}
\usepackage[version=4]{mhchem}
\usepackage{soul}
\usepackage{lettrine}
\usepackage{siunitx}

\author{Djardiel da S. Gomes}
 \affiliation{State University of Campinas, Gleb Wataghin Institute of Physics, Department of Applied Physics, 13083-859, Campinas, São Paulo, Brazil.}
 \alsoaffiliation{University of Bras\'{i}lia, Faculty UnB Planaltina, Materials Science Postgraduate Program, Bras\'{i}lia, Federal District, Brazil.}

\author{Isaac M. Felix}
 \affiliation{Center for Agri-food Science and Technology, Federal University of Campina Grande, 58840-000, Pombal, Paraíba, Brazil.}

\author{Willian F. Radel}
 \affiliation{University of Bras\'{i}lia, Institute of Physics, Physics Postgraduate Program, 70910900, Bras\'{i}lia, Federal District, Brazil.}

\author{Alexandre C. Dias}
 \affiliation{Institute of Physics and International Center of Physics, University of Bras\'ilia, 70910-900, Bras\'ilia, Federal District, Brazil.}
 \alsoaffiliation{University of Bras\'{i}lia, Institute of Physics, Physics Postgraduate Program, 70910900, Bras\'{i}lia, Federal District, Brazil.}

\author{Luiz A. Ribeiro Junior}
 \affiliation{Institute of Physics, University of Bras\'ilia, 70910-900, Bras\'ilia, Federal District, Brazil.}
 \alsoaffiliation{University of Bras\'{i}lia, Institute of Physics, Physics Postgraduate Program, 70910900, Bras\'{i}lia, Federal District, Brazil.}
 \alsoaffiliation{Computational Materials Laboratory, University of Brasília, 70910900, Brasília, Federal District, Brazil.}

\author{Marcelo L. Pereira Junior}
 \affiliation{University of Bras\'{i}lia, College of Technology, Department of Electrical Engineering, 70910-900, Bras\'{i}lia, Federal District, Brazil.}
 \alsoaffiliation{University of Bras\'{i}lia, Institute of Physics, Physics Postgraduate Program, 70910900, Bras\'{i}lia, Federal District, Brazil.}
 \email{marcelo.lopes@unb.br}

\title[]
  {
  Computational Characterization of the Recently Synthesized Pristine and Porous 12-Atom-Wide Armchair Graphene Nanoribbon
  }


\begin{document}

\begin{abstract}
\noindent Recently synthesized Porous 12-Atom-Wide Armchair Graphene Nanoribbons (12-AGNRs) \lbrack\textit{Nano Lett.} 2024, 24, 10718-10723\rbrack exhibit tunable properties through periodic porosity, enabling precise control over their electronic, optical, thermal, and mechanical behavior. This work presents a comprehensive theoretical characterization of pristine and porous 12-AGNRs based on density functional theory (DFT) and molecular dynamics (MD) simulations. DFT calculations reveal substantial electronic modifications, including band gap widening and the emergence of localized states. Analyzed within the Bethe–Salpeter equation (BSE) framework, optical properties highlight strong excitonic effects and significant absorption shifts. Thermal transport simulations indicate a pronounced reduction in conductivity due to enhanced phonon scattering at nanopores. At the same time, MD-based mechanical analysis shows decreased stiffness and strength while maintaining structural integrity. Despite these modifications, porous 12-AGNRs remain mechanically and thermally stable. These findings establish porosity engineering as a powerful strategy for tailoring graphene nanoribbons' functional properties, reinforcing their potential for nanoelectronic, optoelectronic, and thermal management applications.
\end{abstract}

\vspace{0.5cm}

Graphene nanoribbons (GNRs) have garnered significant attention due to their distinctive electronic and structural properties, making them highly suitable for advanced electronic and optoelectronic applications \cite{kumar2023electronic,chen2020graphene,saraswat2021materials,radsar2021graphene,lou2024graphene}. A key advantage of GNRs is their tunable band gap, which can be precisely engineered through width modulation, edge topology control, chemical functionalization, and periodic porosity introduction \cite{luo2022preparation,rizzo2018topological,zhou2020modified}. These strategies have been extensively explored as practical approaches for tailoring electronic properties in low-dimensional materials \cite{qin2024recent,emelianov2024ultrafast,khan2024phase}.  

Recently, Fan \textit{et al.} synthesized porous armchair graphene nanoribbons with a width of \num{12} atoms (Porous 12-AGNR) on an Au($111$) surface using a bottom-up surface-assisted reaction \cite{fan2024bottom}. Their experimental characterization, performed via scanning tunneling microscopy (STM) and scanning tunneling spectroscopy (STS), revealed a substantial band gap increase compared to pristine 12-AGNRs. Complementary Density Functional Theory (DFT) calculations provided initial theoretical insights, yet several key physical properties remain unexplored.  

In this work, we build upon these findings by providing a comprehensive theoretical characterization of pristine and porous 12-AGNRs. We investigate their energetic, thermal, and dynamic stability using advanced DFT-based calculations. Additionally, we refine previously reported electronic properties by employing the hybrid Heyd-Scuseria-Ernzerhof (HSE06) functional \cite{heyd2003hybrid,Heyd2006}, which provides an accurate band gap estimation closer to experimental data. Optical properties are examined within the Bethe-Salpeter equation (BSE) formalism \cite{Salpeter1951}, incorporating significant excitonic effects. We employ fully atomistic classical molecular dynamics (MD) simulations to extend our analysis to larger scales to explore these nanoribbons' thermal transport and mechanical properties.

To systematically evaluate the structural, stability, and electronic properties of these nanoribbons, for the first-principles calculations based on DFT, we used the Spanish Initiative for Electronic Simulations with Thousands of Atoms (SIESTA) code~\cite{soler2002,hohenberg1964,kohn1965,artacho1999}. Structural relaxations and preliminary electronic structure calculations employed the Perdew-Burke-Ernzerhof (PBE) generalized gradient approximation (GGA)~\cite{perdew1996,perdew1997}, widely recognized for its balance between computational efficiency and accurate description of atomic structure. To address the well-known band-gap underestimation of GGA functionals~\cite{xiao2011accurate}, additional calculations were performed using the hybrid HSE06 functional~\cite{heyd2003hybrid,Heyd2006} via the HONPAS (Hefei order-$N$ packages for \textit{ab initio} simulations) package~\cite{Qin2015,Shang2020}, which is entirely based on SIESTA algorithm and numerical methods. We used norm-conserving Troullier-Martins pseudopotentials~\cite{troullier1991,hamann1979} in Kleinman-Bylander form~\cite{kleinman1982}, with a kinetic energy cutoff of 500~Ry and a double-$\zeta$ polarized (DZP) basis set. Brillouin zone (BZ) integration was performed using a Monkhorst-Pack $k$-point grid of $10\times1\times1$~\cite{monkhorst1976}. Structural optimizations included complete relaxation of atomic positions and lattice vectors until residual atomic forces were below 0.001~eV/\r{A}, with energy convergence set at $10^{-5}$~eV. Periodic boundary conditions were applied along the ribbon axis ($x$-direction), while vacuum regions of 35~\r{A} ($y$-direction) and 50~\r{A} ($z$-direction) minimized spurious interactions.

\begin{figure*}[!t]
\includegraphics[width=0.75\linewidth]{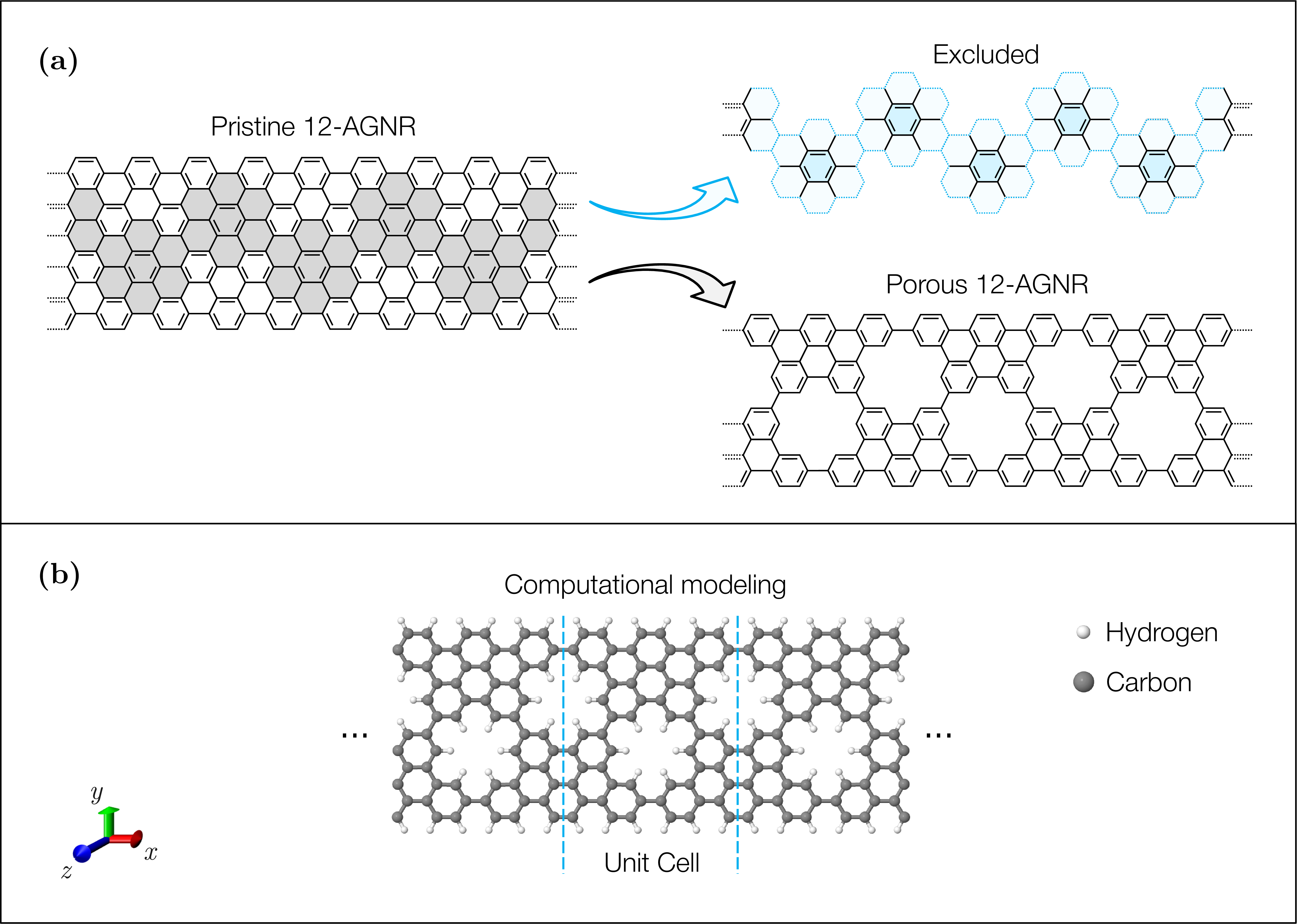}
\caption{Atomic structures of the 12-AGNR models considered in this investigation. (a) Pristine and porous 12-AGNRs, where the porous structure is obtained by removing specific benzene rings and passivating the edges with hydrogen. (b) The computational model used in the simulations highlights the periodic unit cell. Carbon and hydrogen atoms are represented in gray and white, respectively.}
\label{fig:structure}
\end{figure*}

The optimized atomic structures of pristine and porous 12-AGNRs are shown in Fig.~\ref{fig:structure}. Porous 12-AGNRs were generated by selectively removing adjacent benzene rings along the ribbon axis, followed by hydrogen passivation of the exposed carbon atoms. This procedure is illustrated in Fig.~\ref{fig:structure}(a). The final computational model adopted in this study is shown in Fig.~\ref{fig:structure}(b), where the unit cell consists of \num{84} atoms. To ensure a consistent comparison, all properties were also computed for pristine 12-AGNRs using a $3\times1\times1$ supercell, yielding the same number of atoms as the porous structure.

Pristine 12-AGNR exhibits \ce{C-C} bond lengths ranging from 1.38~\r{A} to 1.45~\r{A}, with bonds parallel to the ribbon axis measuring approximately 1.44~\r{A} at the center and 1.38~\r{A} at the edges. In the porous 12-AGNR, bond lengths range from 1.38~\r{A} (at the edges) to 1.50~\r{A} (between consecutive pores), with increased variations around the pore regions. In both cases, \ce{C-H} bonds remain at 1.10~\r{A}, and the structures retain PMMM symmetry (Schoenflies notation: D$_{2h}$-5). The lattice vector along the ribbon axis measures 12.97~\r{A} for pristine 12-AGNR and 13.09~\r{A} for porous 12-AGNR.

Dynamic and thermal stability were further assessed. Phonon dispersion calculations were performed using a $3\times1\times1$ supercell for porous 12-AGNR and a $9\times1\times1$ supercell for pristine 12-AGNR, with an 800~Ry mesh cutoff. Energy and force convergence criteria were set at $10^{-5}$~eV and 0.04~eV/\r{A}, respectively, with the acoustic sum rule enforced at the $\Gamma$ point. The phonon dispersions (Fig.~S1) reveal that both systems are dynamically stable, as no imaginary frequencies appear. While the acoustic branches in pristine 12-AGNR are relatively flat, those in the porous counterpart exhibit increased dispersion due to the additional degrees of freedom introduced by porosity. In both cases, high-frequency modes around $\sim$3200~cm$^{-1}$ correspond to \ce{C-H} bond-length vibrations.

Thermal stability was examined via \textit{ab initio} molecular dynamics (AIMD) simulations at 300~K and 1000~K using a 168-atom supercell ($6\times1\times1$ for 12-AGNR and $2\times1\times1$ for porous 12-AGNR) within the canonical (NVT) ensemble. A Nosé-Hoover thermostat~\cite{evans1985nose} controlled temperature, while a Parrinello-Rahman barostat~\cite{parrinello1981polymorphic} regulated in-plane pressure. As shown in Fig.~S2, both systems maintained structural integrity under thermal conditions, with no significant atomic rearrangement. Additionally, these simulations provided estimates of the specific heat capacity, which remained comparable between pristine and porous 12-AGNRs, indicating that porosity does not significantly affect this parameter. Since these nanoribbons have already been experimentally synthesized, our stability calculations further validate the accuracy and reliability of the computational methods employed in this study.

\begin{figure}[t!]
\includegraphics[width=0.95\linewidth]{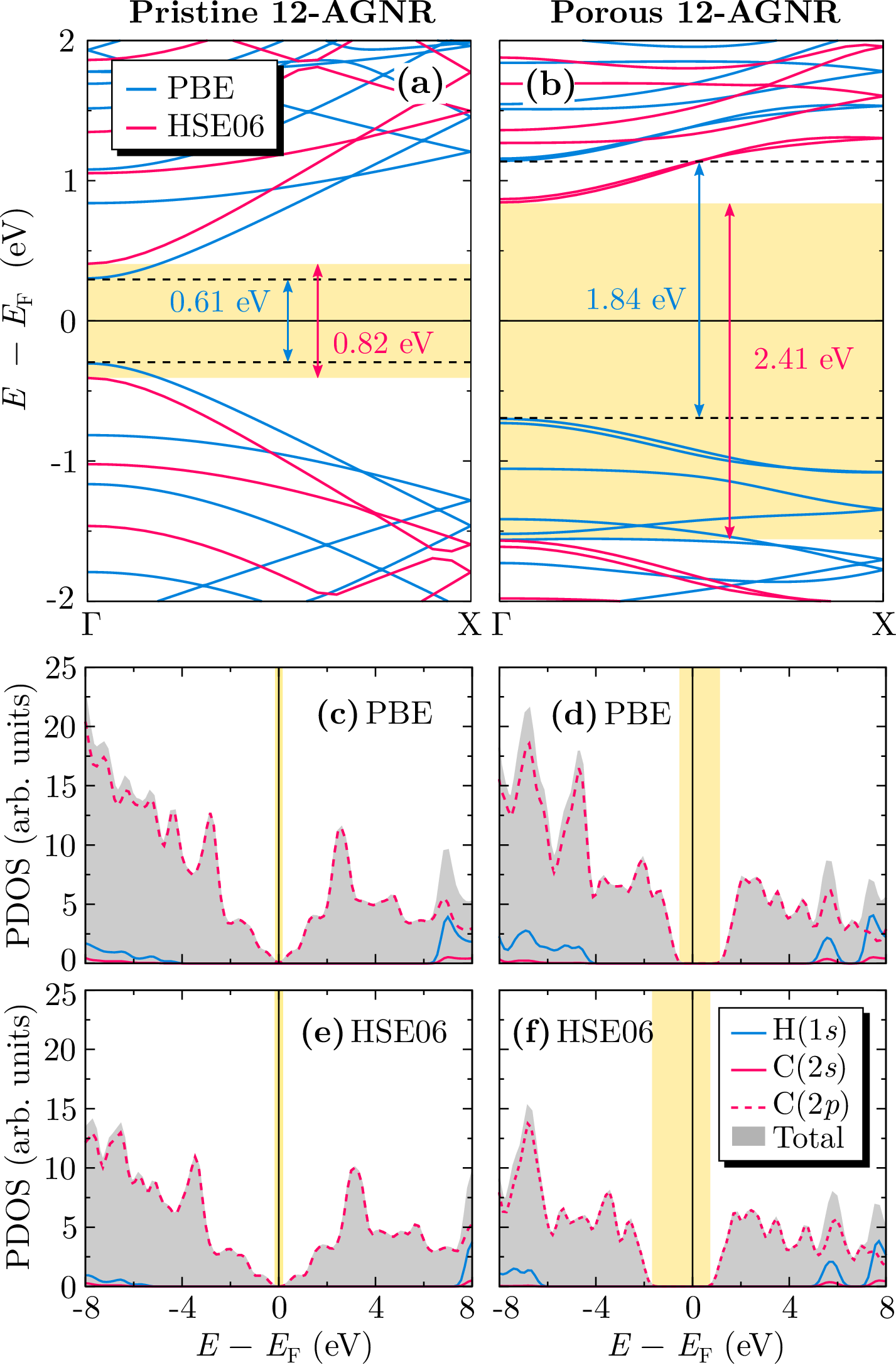}
\caption{Electronic band structures and projected density of states (PDOS) of pristine and porous 12-AGNR nanoribbons. (a, b) Band structures computed using PBE (blue) and HSE06 (red) functionals, with shaded regions highlighting the band gap values. The numerical annotations indicate the energy gaps obtained for each functional. (c-f) PDOS for pristine (c,e) and porous (d,f) 12-AGNR nanoribbons, calculated using PBE (top row) and HSE06 (bottom row) functionals. The contributions of hydrogen (1$s$) and carbon (2$s$ and 2$p$) orbitals are shown separately, with the total density of states represented by the shaded gray region. The yellow-highlighted areas indicate the band gap region.}
\label{fig:band}
\end{figure}

After confirmation of structural and thermal stability, we analyzed the electronic properties of pristine and porous 12-AGNRs. Fig.~\ref{fig:band} presents the electronic band structures and projected density of states (PDOS), computed using PBE and HSE06 functionals.

For pristine 12-AGNR (Fig.~\ref{fig:band}(a)), a direct band gap of 0.61~eV and 0.82~eV was obtained with PBE and HSE06, respectively, with the conduction band minimum (CBM) and valence band maximum (VBM) located at the $\Gamma$ point. These results align with previous theoretical calculations~\cite{fan2024bottom}, where a gap of $\sim$0.6~eV was estimated using GGA-PBE. As expected, HSE06 predicts a larger band gap due to its improved treatment of exchange interactions. Experimental STS measurements reported a band gap of 1.13~eV for pristine 12-AGNRs on Au($111$) substrates.

Introducing periodic porosity (Fig.~\ref{fig:band}(b)) substantially modifies the electronic structure. The band gap remains direct but increases to 1.84~eV (PBE) and 2.41~eV (HSE06), attributed to charge redistribution induced by periodic pore formation. This widening is consistent with experimental reports~\cite{fan2024bottom}, where STS measurements indicated a gap of $\sim$3.3~eV for porous 12-AGNRs on Au($111$). Theoretical estimates from the synthesis study predicted a band gap of 1.9~eV, closely matching our PBE results, whereas our HSE06 calculations better approximate the experimental values. The larger gap observed in the experimental STS data can be attributed to the dielectric screening effect of the Au($111$) substrate and the structural modifications generated by the interaction of the monolayer with the substrate, which modifies the quasiparticle energies and enhances many-body interactions, leading to an increased measured gap~\cite{xu2018interfacial}.

Beyond the band gap expansion, the conduction band in porous 12-AGNR exhibits flatter electronic states relative to its pristine counterpart, suggesting enhanced charge carrier localization due to periodic pore distribution. These features can impact electronic transport by reducing mobility. This effect has been experimentally attributed to interactions between pore edges and the substrate~\cite{fan2024bottom}. Furthermore, theoretical studies indicate that the electronic structure of porous 12-AGNR strongly resembles that of 3-AGNRs, implying that these subunits dominate its electronic behavior.

Hydrogen passivation at the ribbon edges and pore boundaries ensures electronic stabilization, preventing the formation of metallic or highly reactive states from unsaturated carbon atoms. Consequently, the observed electronic features arise intrinsically from the periodic porosity rather than structural defects.

These findings are further supported by the PDOS analysis (Fig.~\ref{fig:band}(c-f)). As expected, the electronic behavior near the Fermi level is predominantly governed by carbon $2p$ orbitals in both valence and conduction bands, reinforcing the $\pi$ character of the electronic states. Notable differences emerge between the two nanoribbon configurations. In pristine 12-AGNRs (Figs.~\ref{fig:band}(c,e)), states near the Fermi level exhibit a relatively continuous distribution, indicative of a homogeneous electronic structure. Conversely, for porous 12-AGNRs (Figs.~\ref{fig:band}(d,f)), a pronounced increase in the separation of electronic states is observed, consistent with the previously discussed band gap widening. Additionally, modifications in the conduction band density of states further support localized electronic states induced by periodic porosity.

The substantial band gap modulation caused by porosity suggests a significant impact on the optical properties of these nanoribbons, particularly in excitonic effects and absorption spectra. To capture these many-body interactions, we employ the BSE \cite{Salpeter1951} formalism to describe electron-hole interactions. It provides a detailed characterization of excitonic states and their contributions to the optical response.

Excitonic and optical properties were computed using the WanTiBEXOS code \cite{Dias_108636_2023} within the independent-particle approximation (IPA) and the BSE framework. A $25\times1\times1$ \textbf{k}-mesh was adopted, considering \num{14} conduction and valence bands for pristine 12-AGNRs and \num{10} conduction and \num{8} valence bands for porous 12-AGNRs. However, direct interface compatibility with the SIESTA/HONPAS Hamiltonian is not publicly available but can be provided upon request. A detailed description of the BSE implementation, including the excitonic Hamiltonian formulation and computational parameters, is provided in the Supporting Information.

\begin{figure}[t!]
\begin{center}
\includegraphics[width=0.95\linewidth]{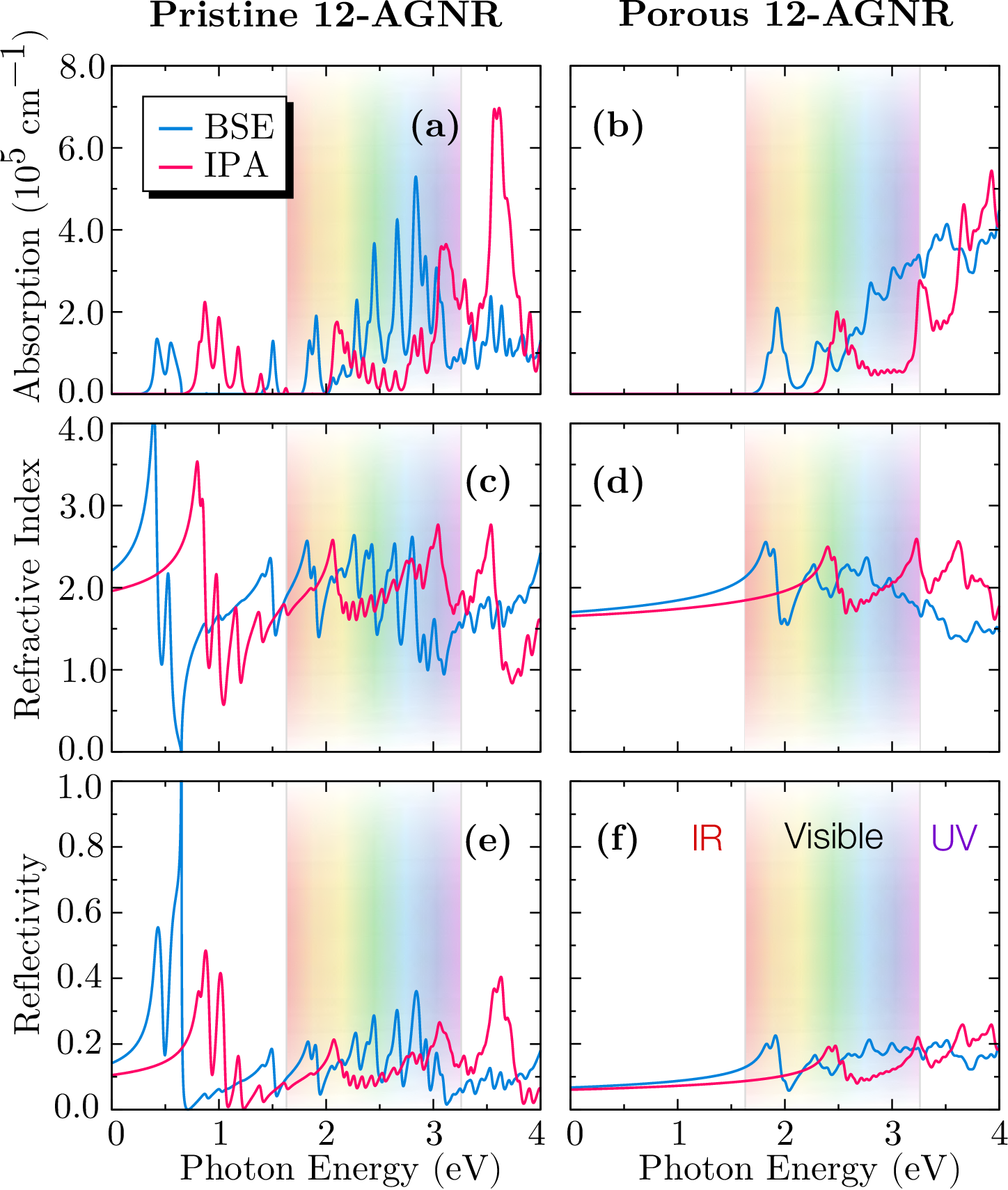}
\caption{Optical properties of pristine (left) and porous (right) 12-AGNRs. (a,b) Optical absorption coefficient, (c,d) refractive index, and (e,f) reflectivity as a function of photon energy.}
\label{fig:optic}
\end{center}
\end{figure}

Fig.~\ref{fig:optic} presents the linear optical response in terms of reflectivity, refractive index, and absorption coefficient, obtained within the IPA (pink curves) and BSE (blue curves) for pristine (left panels) and porous (right panels) 12-AGNRs. A pronounced redshift in the optical band gap is observed in panels (a) and (b), leading to estimated exciton binding energies of \SI{400}{\milli\electronvolt} for pristine and \SI{787}{\milli\electronvolt} for porous 12-AGNRs. These values exceed those reported for two-dimensional carbon allotropes~\cite{Dias_8572_2024}, a consequence of the enhanced quantum confinement in quasi-1D systems. Both structures exhibit absorption across the infrared (IR), visible, and ultraviolet (UV) regions, with higher absorption coefficients at shorter wavelengths. 

Pristine 12-AGNRs exhibit a stronger linear optical response than their porous counterparts, with a higher refractive index and reflectivity. Notably, total reflection is achieved for photons in the infrared spectrum, with partial reflection extending into the visible and ultraviolet regions. In contrast, porous 12-AGNRs display minimal reflection, limited to a narrow range in the visible and UV regions. These findings highlight the influence of periodic porosity on optical properties, reinforcing its role in tailoring the light-matter interaction in graphene-based nanostructures.

Large-scale atomic systems were considered to characterize the thermal transport and mechanical response of pristine and porous 12-AGNRs, as a fully quantum-mechanical treatment would be computationally prohibitive. To accurately capture long-wavelength phonon transport and mechanical response, supercells containing between \num{336} and \num{13440} atoms were employed, reaching lengths up to \SI{200}{\nm}. Given these system sizes, we utilized classical MD simulations, which enable the investigation of atomic-scale dynamics at spatial and temporal scales inaccessible to first-principles methods. In classical MD, atoms are modeled as classical particles governed by Newton’s equations:
\begin{equation}
    m_i \frac{d^2 \mathbf{r}_i}{dt^2} = - \frac{\partial E}{\partial \mathbf{r}_i},
\end{equation}
where $m_i$ and $\mathbf{r}_i$ denote the mass and position of atom $i$, respectively, and $E$ is the total potential energy. The accuracy of the computed properties depends directly on the choice of interatomic potential. Here, atomic interactions were described using the second-generation Reactive Empirical Bond Order (REBO) potential~\cite{Brenner2002}, widely validated for carbon-based nanostructures~\cite{wei2022,felix2024,mortazavi2014,kim2021,jiang2010,chien2012}. All MD simulations were performed using the Large-scale Atomic/Molecular Massively Parallel Simulator (LAMMPS)~\cite{Thompson2022}, with equations of motion integrated via the velocity-Verlet algorithm \cite{verlet1967computer}.

Thermal transport properties were investigated using the reverse non-equilibrium molecular dynamics (RNEMD) method proposed by Müller-Plathe~\cite{Muller1997}. Periodic boundary conditions were imposed along the longitudinal ($x$-axis) direction, and nanoribbons up to \SI{200}{\nm} long (\num{13440} atoms) were partitioned into \num{20} slabs. Heat flux was induced by exchanging kinetic energy between high-velocity atoms at the extremities and low-velocity atoms at the center. After equilibration, a stable temperature gradient was established, enabling the calculation of the lattice thermal conductivity $\kappa(L_x)$ as:
\begin{equation}\label{eq:RNEMD}
    \kappa (L_x) = - \frac{1}{\nabla_x T} \left[ \frac{\sum_{\text{swaps}} \Delta K}{2 A \Delta t} \right],
\end{equation}
where $A$ is the cross-sectional area, given by the product of the ribbon width ($\sim$\SI{1.5}{\nm}) and thickness ($h=$ \SI{0.335}{\nm}). The thermal conductivity follows a ballistic-to-diffusive length dependence, modeled as:
\begin{equation}\label{eq:kappa-length}
    \frac{1}{\kappa(L_x)} = \frac{1}{\kappa_\infty}\left(1+\frac{\Lambda}{L_x}\right),
\end{equation}
where $\kappa_\infty$ is the intrinsic thermal conductivity and $\Lambda$ the effective phonon mean free path. Simulations were performed with a timestep of \SI{0.5}{\fs}, with kinetic energy swaps every \num{500} timesteps, over a total duration of \SI{20}{\ns}.

\begin{figure}[b!]
\begin{center}
\includegraphics[width=0.99\linewidth]{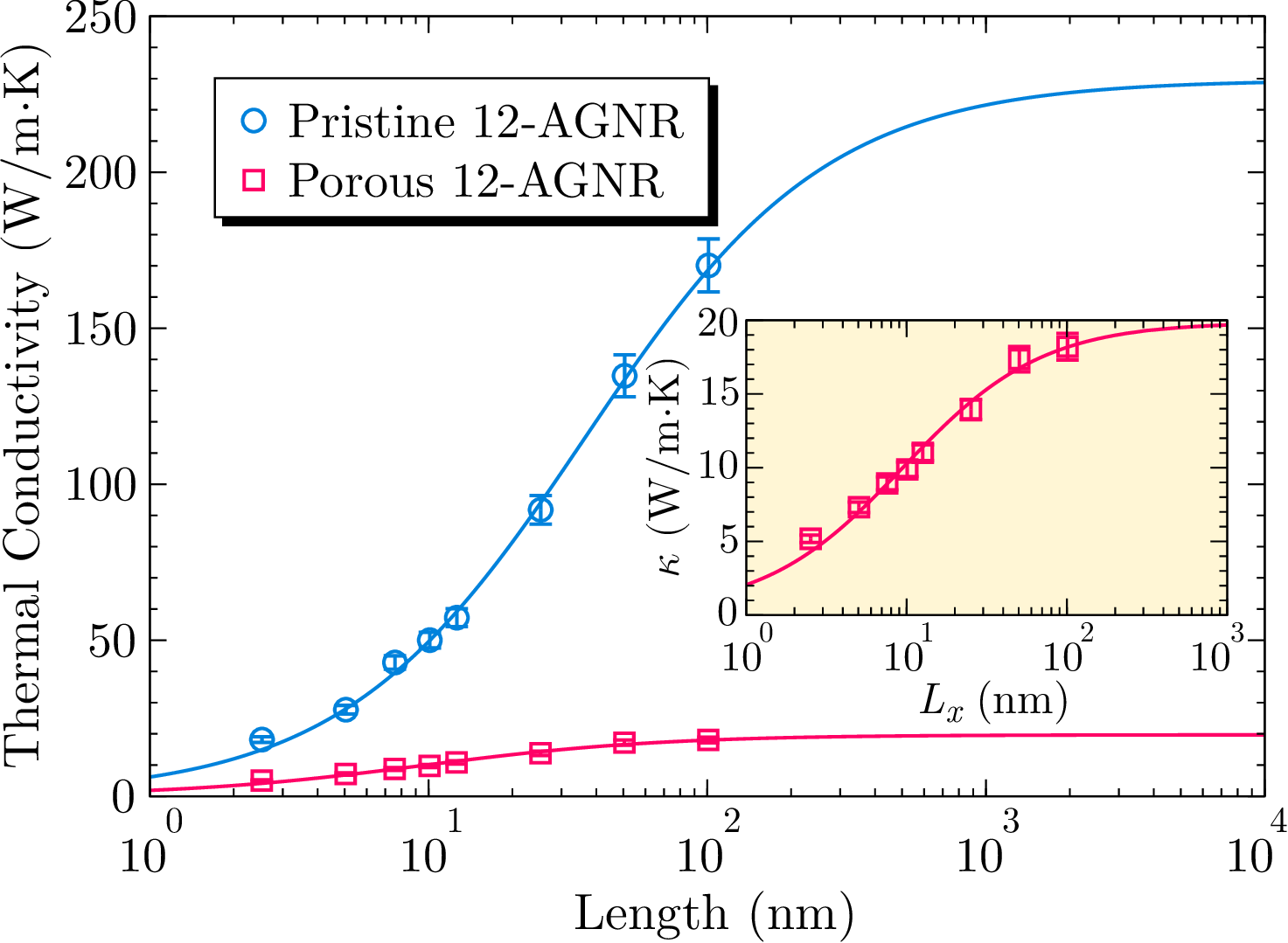}
\caption{Thermal conductivity of pristine and porous 12-AGNRs as a function of sample length. Data points correspond to RNEMD simulations, while solid lines are fitted using Eq.~(\ref{eq:kappa-length}).}
\label{fig:kappa}
\end{center}
\end{figure}

Fig.~\ref{fig:kappa} presents the dependence of thermal conductivity on nanoribbon length, with pristine 12-AGNRs shown for comparison. Symbols represent values computed via Eq.~\eqref{eq:RNEMD}, while solid lines correspond to least-squares fits using Eq.~\eqref{eq:kappa-length}. Independent simulations with varying initial atomic velocities were conducted to estimate uncertainties, yielding an uncertainty below \SI{5}{\percent} in all cases. Fluctuations in heat flux and temperature gradient were also considered, confirming a similar uncertainty level. The agreement between fitted curves and computed data highlights the predictive capability of Eq.~\eqref{eq:kappa-length}, enabling the estimation of intrinsic thermal conductivity from relatively short simulations.

The intrinsic thermal conductivity of pristine 12-AGNRs was $230 \pm 22$ W/m$\cdot$K, in agreement with previous reports~\cite{kim2021,liu2018}. As expected, porosity significantly reduces the thermal conductivity, yielding an estimated $19.7 \pm 1.7$ W/m$\cdot$K. The dependence of thermal transport on pore density, size, and spatial distribution is well documented~\cite{yang2015,yarifard2016,hu2018,yousefi2020}, with our results indicating a $\sim$\SI{90}{\percent} reduction in $\kappa$ compared to pristine 12-AGNRs. This decline is comparable to that observed in GNRs with isotopic doping~\cite{jiang2010}, chemisorption functionalization~\cite{chien2012}, and Stone-Thrower-Wales defects~\cite{ng2012}, though less pronounced than in kirigami-engineered GNRs, where $\kappa$ is reduced by approximately two orders of magnitude~\cite{wei2016}.

Beyond thermal transport, structural modifications such as porosity and edge morphology also influence mechanical properties. We performed uniaxial tensile deformation simulations along the longitudinal direction to assess these effects. Given the computational demands of such analyses, shorter nanoribbons of approximately \SI{50}{\nm} (\num{3276} atoms) were used. For statistical robustness, ten independent simulations were conducted for each configuration at room temperature, varying initial atomic velocity distributions. Each system was equilibrated for \SI{100}{\ps} at \SI{300}{\kelvin} before applying a continuous uniaxial deformation over \SI{5}{\ns} at a constant strain rate of \SI{E-4}{\per\ps}. The velocity-Verlet algorithm was employed for numerical integration, with a timestep of \SI{0.1}{\fs}.

Atomic-level stress components were computed via the virial theorem and normalized by the inverse of the effective volume $V$:
\begin{equation}
\sigma_{\alpha \beta} = -\frac{1}{V} \left[ \sum_i m_i v_{i\alpha} v_{i\beta} + \sum_{i,j} f_{ij\alpha} r_{ij\beta} \right],
\end{equation}
where $m_i$ is the mass of atom $i$, $v_{i\alpha}$ and $v_{i\beta}$ are velocity components, $f_{ij\alpha}$ is the interatomic force between atoms $i$ and $j$, and $r_{ij\beta}$ is the displacement between them. The first term accounts for kinetic contributions, while the second represents interatomic interactions \cite{rodrigues2025machine,de2023nanomechanical}. The effective volume is computed as $V = L_x \times W \times h$, where $L_x$ is the ribbon length, $W$ is the previously mentioned width of approximately \SI{1.5}{\nm}, and $h$ is the thickness.

The uniaxial strain was defined as $\varepsilon = \Delta L / L_0$, where $L_0$ is the initial length and $\Delta L$ is the corresponding change upon deformation. In the elastic regime, stress follows Hooke's law, $\sigma = Y_M \varepsilon$, where $Y_M$ is Young's modulus. Stress-strain curves were computed consistently with the ribbon thickness (\SI{0.335}{\nm}) adopted in the thermal transport analysis, ensuring direct comparability between thermal and mechanical properties.

\begin{figure}[t!]
\begin{center}
\includegraphics[width=0.95\linewidth]{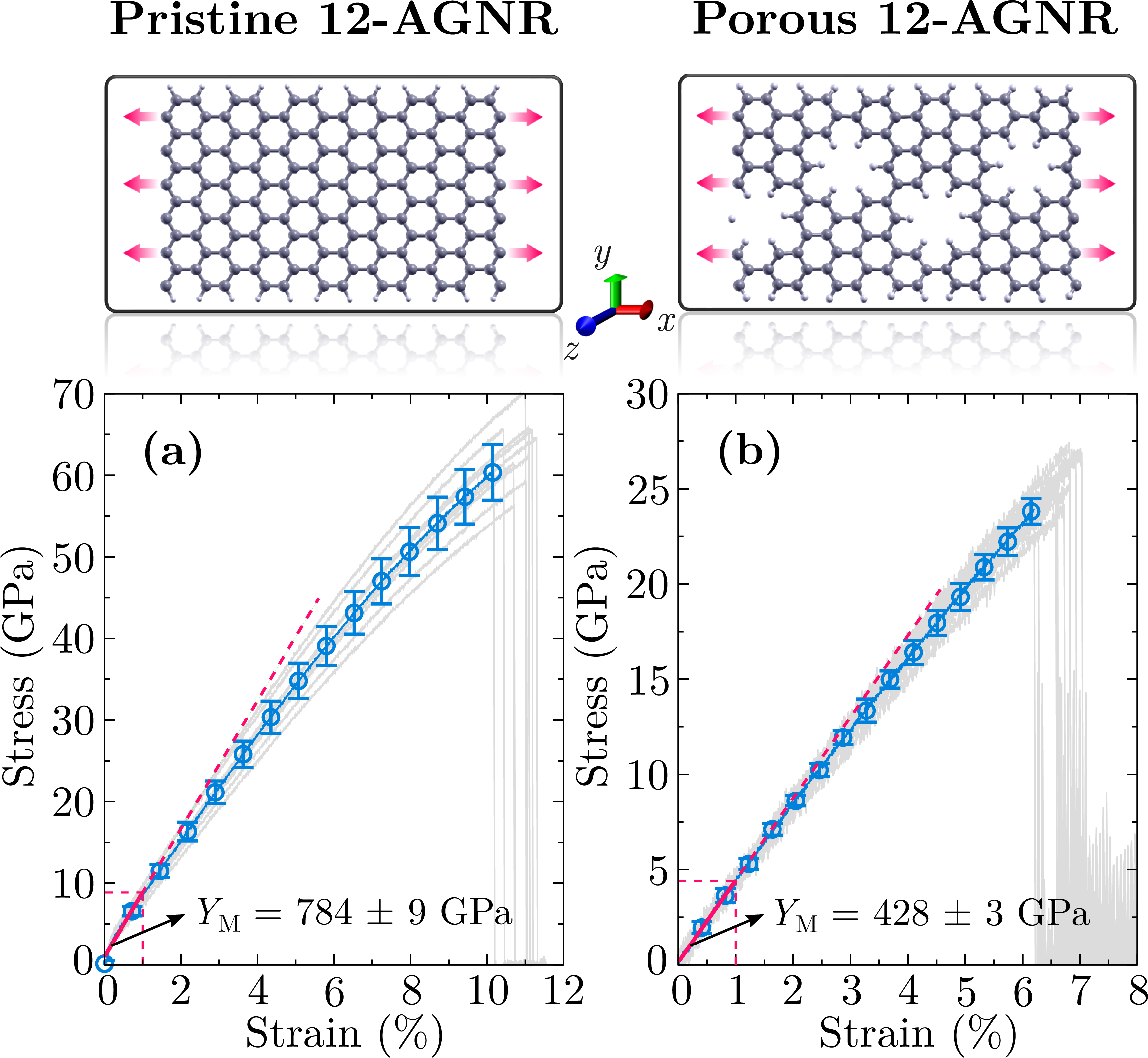}
\caption{Stress-strain response of pristine (a) and porous (b) 12-AGNR nanoribbons under uniaxial tensile deformation.}
\label{fig:stress-strain}
\end{center}
\end{figure}

Fig.~\ref{fig:stress-strain} presents the stress-strain behavior of pristine and porous 12-AGNRs under uniaxial tensile loading. In the elastic regime, stress increases linearly with strain, allowing for the extraction of $Y_M$ by fitting the data to Hooke's law. Pristine 12-AGNRs exhibit a $Y_M$ of 784 $\pm$ 9 GPa, while porous 12-AGNRs display a 428 $\pm$ 3 GPa reduced modulus. These values align well with DFT calculations, which yield $Y_M = 864$ GPa and $Y_M = 494$ GPa for pristine and porous 12-AGNRs, respectively. A slight overestimation in DFT results is expected, as finite-temperature effects inherent to MD simulations are absent in static DFT calculations, reinforcing the reliability of MD for mechanical property predictions. The observed $\sim$\SI{45}{\percent} reduction in Young's modulus reflects the weakening of the carbon network due to periodic pore formation, which disrupts the delocalized $\pi$-bonding responsible for the high stiffness of graphene-based materials. 

In addition to reducing stiffness, porosity also weakens the mechanical strength of nanoribbons, leading to lower critical strain and stress values. For pristine 12-AGNRs, the critical stress and strain are $\sigma_C = 63\pm 4$ GPa and $\varepsilon_C = 10.9\pm 0.4\%$, respectively. In porous 12-AGNRs, these values drop to $\sigma_C = 25\pm 1$ GPa and $\varepsilon_C = 6.7\pm 0.3\%$, demonstrating the direct impact of periodic voids on fracture resistance. The fracture mechanism is brittle in all cases, with failure occurring abruptly and no significant plastic deformation. These results indicate that periodic porosity is an effective strategy for modulating the mechanical response of graphene nanoribbons while maintaining structural integrity.

Beyond mechanical properties, periodic porosity enables a highly tunable platform where electronic, optical, and thermal properties can be systematically engineered.

In summary, we have conducted a comprehensive theoretical investigation of pristine and porous 12-AGNRs, a recently synthesized graphene nanoribbon system, analyzing their structural, electronic, optical, thermal, and mechanical properties. First-principles calculations revealed that periodic porosity significantly modifies the electronic structure, widening the band gap and introducing localized states. Optical properties demonstrated strong excitonic effects, resulting in a significant red shift in the optical band gap, emphasizing the role of many-body interactions. Thermal transport simulations showed a substantial reduction in conductivity due to phonon scattering at nanopores, while mechanical analysis confirmed a decrease in stiffness and strength with retained structural stability. These findings establish porosity engineering as a versatile tool for tailoring the functional properties of graphene nanoribbons, reinforcing their potential for nanoelectronic, optoelectronic, and thermal management applications.

\begin{acknowledgement}
This work received partial support from the Brazilian Coordination for the Improvement of Higher Education Personnel (CAPES), the National Council for Scientific and Technological Development (CNPq), and the Research Support Foundation of the Federal District (FAPDF).  

D.G.S. acknowledges financial support from CAPES under grant No. 88887.102348/2025-00 (Finance Code 001).

W.F.R. acknowledges financial support from CAPES under grant No. 88887.840299/2023-00 (Finance Code 001).

L.A.R.J. acknowledges financial support from FAPDF (grants 00193.00001808/2022-71 and 00193-00001857/2023-95), FAPDF-PRONEM (grant 00193.00001247/2021-20), PDPG-FAPDF-CAPES Centro-Oeste (grant 00193-00000867/2024-94), and CNPq (grants 350176/2022-1 and 167745/2023-9).

A.C.D. acknowledges financial support from FAPDF (grants 00193-00000867/2024-94, 00193-00001817/2023-43, and 00193-00002073/2023-84) and CNPq (grants 408144/2022-0, 444069/2024-0, and 444431/2024-1). Computational resources were provided by the National High-Performance Computing Center in São Paulo (CENAPAD-SP, UNICAMP, project proj960) and the High-Performance Computing Center (NACAD, Lobo Carneiro Supercomputer, UFRJ, project a22002).  

M.L.P.J. acknowledges financial support from FAPDF (grant 00193-00001807/2023-16) and CNPq (grant 444921/2024-9). He also acknowledges computational support from CENAPAD-SP (project 897) and NACAD (project 133).
\end{acknowledgement}

\begin{suppinfo}
The Supporting Information includes phonon dispersion calculations confirming the dynamic stability of both pristine and porous 12-AGNRs. Additionally, AIMD simulations at 300~K and 1000~K validate their thermal stability, demonstrating structural integrity under ambient and elevated temperatures. Further details on the BSE formalism are provided, including computational parameters and methodologies used for excitonic state calculations.
\end{suppinfo}

\bibliography{references}

\end{document}